\documentclass{JHEP3}
\usepackage{graphicx}
\usepackage{epsfig}
\newcommand{\gev}{~\mathrm{GeV}}
\newcommand{\tev}{~\mathrm{TeV}}
\newcommand{\hnn}{h \to \chi\chi}
\newcommand{\brhnn}{BR$(h\to \chi\chi)$~}

\title{Higgs decays in supersymmetric models with light neutralinos}
\author{Carlos E. Yaguna \\ Department of Physics and Astronomy, UCLA, Portola
Plaza, Los Angeles, CA 90095, USA\\ \email{yaguna@physics.ucla.edu}}
\abstract{In the Minimal Supersymmetric Standard Model, neutralinos lighter than $50\gev$ are compatible with
  all accelerator, precision, and cosmological bounds. Such
  neutralinos might constitute a relevant decay channel for the  Higgs
  boson, modifying its expected signatures at hadron colliders. We study the
  branching ratio $\hnn$ and determine the region in the supersymmetric
  parameter space where it is sizable. We have found that, in fact, the Higgs
  may dominantly decay into neutralino pairs. Besides, as a result of this new channel, the
  branching ratio into visible modes, such as $h\to \gamma\gamma$, gets suppressed.}

\begin{document}
\section{Motivation}
The Large Hadron Collider (LHC), now in  its final stages, may  soon provide  evidence of  new
physics at the TeV scale. First of all, it should  discover the Higgs boson, the
only missing particle in the Standard Model. Indeed, current
data from direct searches \cite{Barate:2003sz} and electroweak precision
measurements \cite{:2005em} suggests a
Higgs mass in the range $114\gev<m_h<250\gev$ --well within the reach of the
LHC. Once produced, the Higgs boson will instantaneously decay into lighter particles. Most
analysis have assumed that its decay products are Standard Model particles: $bb$,
$\tau\tau$, $\gamma \gamma$, etc. That is not necessarily the case, however,  if new
physics, such as low energy
supersymmetry, exists at the TeV scale.

Low energy supersymmetry is by far the best motivated
scenario for physics beyond the Standard Model. The minimal supersymmetric
extension of the Standard Model, known as the  MSSM, solves the gauge
hierarchy problem, achieves the unification of
gauge couplings, explains the origin of electroweak symmetry breaking, and  includes a natural dark matter candidate: the lightest
neutralino.

In the MSSM the mass of the lightest
neutralino is not constrained by present experiments. The often cited bound
$m_\chi> 50\gev$ comes from the LEP limit on the chargino mass -$M_2,
|\mu|>103\gev$ \cite{Yao:2006px}- and assumes the GUT relation among gaugino masses, which leads to
$M_1\approx 0.5 M_2$ at the electroweak scale. This relation, however, does
not necessarily hold, not even in models with unification of gauge couplings
at high scales. And good theoretical as well as phenomenological arguments in favor of models with
non-universal gaugino masses have been put forward in the literature \cite{Hill:1983xh}. If the GUT relation is ignored and $M_1$ is considered as an independent
parameter, the neutralino mass can be much smaller than $50\gev$.

Accelerator constraints on  light neutralinos are not strong
if the neutralino is bino-like. The
invisible width of the Z boson, for instance, usually constraints the
existence of light particles. For a pure bino, however, the tree level
coupling to the Z boson vanishes and no bounds can be derived.
Indeed, according to the Review of Particle Physics \cite{Yao:2006px}: ``a bino of mass $0.1\mathrm{MeV}$ is not excluded by collider
experiments''. In a recent study,  Dreiner et al \cite{Dreiner:2007fw}
concluded, along similar lines, that  in the MSSM
the mass of the lightest neutralino is unconstrained.

Within the standard cosmological model,  light neutralinos may also explain the dark matter of the
Universe \cite{Spergel:2006hy}. In a number  of papers \cite{Bottino:2002ry},
it has been shown
that neutralinos with masses above $6\gev$ are compatible with the
observed dark matter density as well as   with
current bounds from direct and indirect dark matter searches. Other scenarios
where light neutralinos may play a role as dark matter candidates are  non-standard
cosmological models, as recently pointed out by Gelmini et al
\cite{Gelmini:2006mr}. They found supersymmetric models with  neutralino
masses as low as $1\gev$ that are consistent with present bounds from
accelerator and dark matter searches.

Supersymmetric models with light neutralinos, therefore, are viable and
well-motivated scenarios for physics beyond the standard model. A remarkable feature of these models  is the possibility that the Higgs
boson decays into a neutralino pair ($h\to \chi\chi$) with a sizable branching
ratio. Such invisible decay would have important implications for Higgs
searches at the Tevatron and at the LHC. The detectability of a hypothetical
Higgs boson that decays invisibly has been considered in several studies
\cite{Frederiksen:1994me,Godbole:2003it,Davoudiasl:2004aj}. And
they have all mentioned the MSSM with light neutralinos as one of its possible
realizations. A detailed analysis of the process $\hnn$ in the MSSM, however,
has not been published. This paper presents such an analysis. As a
first step, the supersymetric parameters that determine the value of
\brhnn are identified. Next, we compare $\hnn$ with $h\to bb$ and show that
neutralinos may indeed constitute the dominant decay mode of the Higgs
boson. The behavior of   \brhnn is then studied as a function  of $M_1$, $\tan\beta$, and
$\mu$. Finally, we point out that by increasing the Higgs total decay width,
the decay  $\hnn$  suppresses the Higgs branching ratio into visible channels,  such as $h\to \gamma \gamma$.

\section{Phenomenology}

The scalar sector of the MSSM contains five physical degrees of freedom: two
neutral scalar fields ($h^0, H^0$), one neutral pseudo scalar field ($A$), and two
charged  scalar fields ($H^\pm$). In the decoupling limit, when $m_A\gg m_Z$,
the only light Higgs boson is $h^0$ and its coupling to the gauge bosons tend
to those of the Standard Model Higgs boson. We will henceforth refer to $h^0$
as the Higgs boson and denote it simply by $h$. A crucial difference between
the Standard Model and the MSSM is that the Higgs boson is
necessarily light in the MSSM. At tree level its mass already contradicts the current
bound set by LEP ($m_h>114.4\gev$). Higher order radiative corrections to the Higgs
spectrum, however, are important and increase the theoretical
prediction of $m_h$ \cite{Reina:2005ae}. Yet, the Higgs mass  can hardly be above $135\gev$.

The lightest neutralino of the MSSM is a linear superposition of the bino
($\tilde B$), the wino ($\tilde W$), and the two Higgsinos ($H_1^0, H_2^0$):
\begin{equation}
\chi=a_1 \tilde B+a_2\tilde W+a_3 \tilde H_1^0+a_4 \tilde H_2^0\,.
\end{equation}
Due to the structure of the neutralino mass matrix, the neutralino is
bino-like if $M_1\ll M_2,\mu$, wino-like if $M_2\ll M_1,\mu$, and
higgsino-like if $\mu\ll M_1,M_2$. Since the  LEP data put a strong constraint on the
chargino mass -$M_2, \mu>103\gev$-, light neutralinos, $m_\chi<50\gev$, are
dominantly bino-like and have $m_\chi\sim M_1$. Small wino and higgsino
components, however, are not ruled out, and they usually  play an important
phenomenological role. In fact, since the Higgs boson coupling to neutralinos
vanishes when the neutralino is a pure gaugino or a pure higgsino \cite{Djouadi:2005gj},
it is the small bino-higgsino mixing that makes the process $\hnn$ possible at all. 

A problem common to all
phenomenological studies of the MSSM is the vastness of the supersymmetric parameter space. Fortunately, only few parameters are
relevant for the computation of the $\hnn$ branching ratio. We have found that \brhnn is mainly
determined by $\tan\beta$, $\mu$, and $M_1$. At the end, these three
parameters control the all-important bino-higgsino mixture in the lightest
neutralino. In addition, $M_1\sim m_\chi$ also affects the phase space available to
the decaying Higgs. For simplicity, all other
parameters will be  given a specific value throughout this paper. We set $m_A$, $M_2$, $M_3$ and all squark and
slepton masses to $2\tev$; all trilinear couplings are zero except for
the stop coupling which is set to $3\tev$. The effect of these parameters is
 not significant on
\brhnn but it could be important on specific accelerator or precision
bounds. The Higgs mass, for instance, is rather sensitive to the value of the
stop trilinear coupling. On the supersymmetric models  we impose the following
bounds: $m_h>114.4\gev$ \cite{Barate:2003sz},
$m_{\chi^\pm}>103\gev$ \cite{Yao:2006px}, $-15<(g-2)_\mu\times 10^{10}<67$ \cite{Davier:2007ua}, and $2.83<BR(b\to
s\gamma)\times 10^4< 4.27$ \cite{Barberio:2006bi}. For our calculations we use the FeynHiggs
program \cite{Hahn:2006np} which  computes, among others, the masses and mixings of the
Higgs bosons in the MSSM at the two-loop level. In the next section we will
study \brhnn as a function of $\mu$, $\tan\beta$, and $M_1$.

\EPSFIGURE[t]{brtb,scale=0.4}{\brhnn (solid line) and BR($h\to bb$)
  as a function of $\tan\beta$ for $\mu=200\gev$ and
  $M_1=35\gev$.\label{fig:brtb}}

\section{Results}

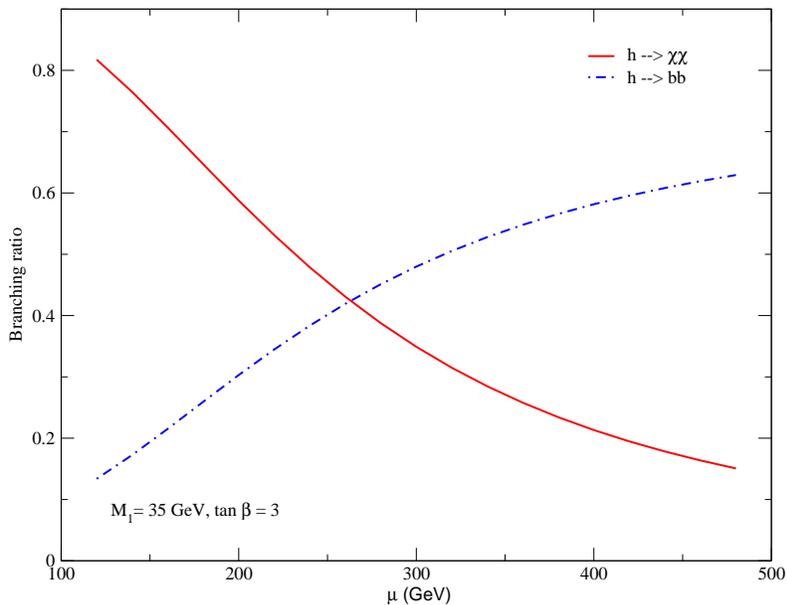
\EPSFIGURE[t]{brmu,scale=0.4}{\brhnn (solid line) and BR($h\to bb$)
  as a function of $\mu$ for $\tan\beta=3$ and $M_1=35\gev$\label{fig:brmu}}

Most analysis of Higgs decays in the MSSM assume that all superpartners are
heavy enough to prevent the decay of the Higgs into pairs of supersymmetric
particles, so that only decays into Standard Model final states are possible. In
that case the dominant decay channel is $bb$, typically  accounting for more
than $80\%$ of the decays \cite{Reina:2005ae}. We will see that a different  picture
emerges  if the Higgs can decay into  neutralino pairs.

Figure \ref{fig:brtb} compares the Higgs branching ratio into neutralinos with
its branching ratio into $b$ quarks  as a function of
$\tan\beta$ for $M_1=35\gev$ and $\mu=200\gev$. Note that these two channels
account for the lion's share of the decays. \brhnn is seen to decrease with $\tan\beta$
whereas BR($h\to bb$) increases with it. At low $\tan\beta$, the decay into neutralinos  dominates,
reaching \brhnn$\sim 70\%$. Around $\tan\beta= 5$ both
decays give similar contributions ($\sim 40\%$) and from then on the decay into $b$ quarks dominates. Yet, \brhnn remains non-negligible ($> 10\%$) all
the way up to $\tan\beta\sim 25$.

In figure \ref{fig:brmu} we compare the same two decays but now  as a function
of $\mu$ for $M_1=35\gev$ and
$\tan\beta=3$. \brhnn is a decreasing function of  $\mu$ whereas BR($h\to bb$)
is an increasing one. For small values of $\mu$, $\hnn$ is the dominant decay mode,
accounting for up to  $80\%$ of the decays.  The two
branching ratios become equal, $\sim 40\%$, around $\mu=250\gev$ and from then on $h\to bb$
dominates. At $\mu=400\gev$, \brhnn has decreased to $20\%$ whereas BR($h\to
bb$) has increased to about $60\%$. 

Thus, the decay $\hnn$ is not a small effect, it may be the dominant decay
mode of the Higgs boson. As we have seen \brhnn is larger when the higgsino component in the
lightest neutralino is larger, that is for low $\tan\beta$ and small $\mu$, but
it remains significant within a wider range.

\EPSFIGURE[t]{mu,scale=0.45}{BR($h\to \chi \chi$) as a function of the
  neutralino mass for $\tan\beta=3$ and different values of $\mu$.\label{fig:mu}}

We proceed now to study  the dependence of \brhnn with the other relevant parameter,
$M_1$, or equivalently $m_\chi$. Because we already know that the decay into neutralinos is important for low
$\tan\beta$ and small $\mu$ we will concentrate on that region of the parameter
space. Figure \ref{fig:mu} shows \brhnn as a function of the neutralino mass for
$\tan\beta=3$ and different values of $\mu$. Due to the reduced phase space,
\brhnn is suppressed when the neutralino mass is close to $m_h/2$. For $m_\chi< 40\gev$,
\brhnn varies  mildly with  $m_\chi$. The dependence with $\mu$, on the
other hand, is very
strong. At small values of $\mu$, when  the bino-higgsino mixing is larger, the
branching ratio may reach $80\%$. For $\mu=300\gev$, not a
small value,  \brhnn gets to  $40\%$ and it reaches
almost $20\%$ for $\mu=500\gev$.

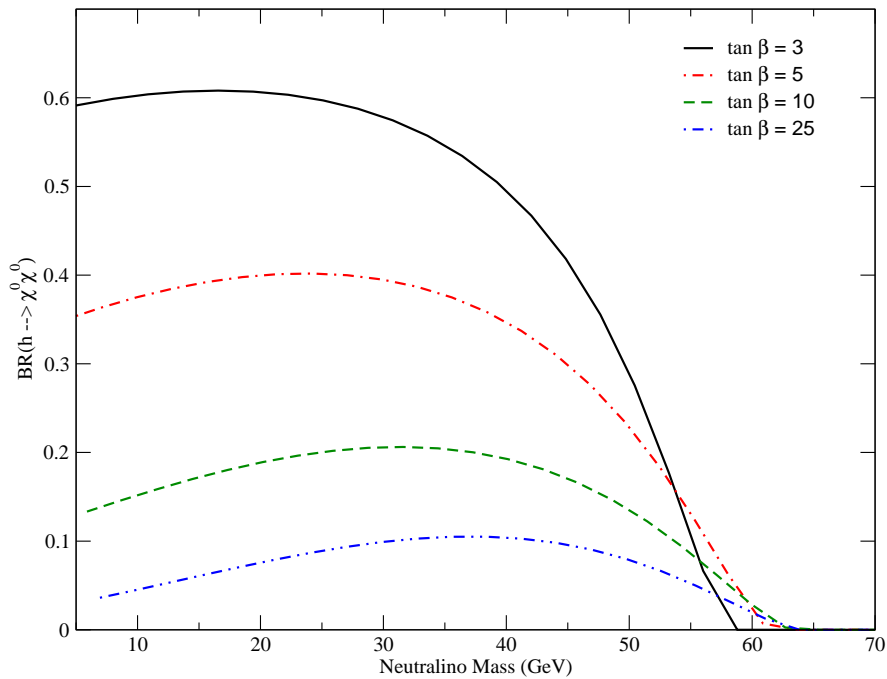
\EPSFIGURE[t]{tanb,scale=0.45}{BR($h\to \chi \chi$) as a function of the
  neutralino mass for $\mu=200\gev$ and several values of $\tan\beta$.\label{fig:tanb}}

In figure \ref{fig:tanb} we illustrate the dependence of \brhnn with the
neutralino mass for different values of $\tan\beta$ and $\mu=200\gev$. As
expected, \brhnn is larger for smaller values of $\tan\beta$ and it goes to
zero as $m_\chi$ approaches $m_h/2$. Since the Higgs mass depends on
$\tan\beta$, \brhnn vanishes at slightly different neutralino masses, as
observed in the figure. As before, the dependence with the neutralino mass is
particularly important for neutralino masses larger than about $40\gev$. For $\tan\beta=3$ \brhnn reaches a
maximum value of $60\%$ that decreases to $10\%$ for $\tan\beta=25$.

Since neutralinos are stable and neutral, the process $\hnn$  is an example of
the so-called invisible decays. Generic models with an invisibly decaying Higgs boson
have been considered before
\cite{Frederiksen:1994me,Godbole:2003it,Davoudiasl:2004aj}. We
have  shown that the MSSM with light neutralinos is indeed one of its
possible realizations.

Discovering a Higgs that decays invisibly would certainly be
challenging, but it is within the capabilities of the LHC. The production of
the Higgs in association with  a $Z$ boson
seems to be the most promising avenue,  providing a clean signal in the
dilepton plus missing energy channel.  Godbole et al \cite{Godbole:2003it}, for instance, concluded that,
for $m_h=120\gev$ and $100~$fb$^{-1}$ of integrated luminosity,
 invisible branching ratios larger than $\sim 0.42$  can be probed at
 $5\sigma$. In a recent study Davoudias et al \cite{Davoudiasl:2004aj}
 considered a $100\%$ invisibly decaying Higgs and found that the $Z+h$
 channel can provide a discovery with  $10~$fb$^{-1}$ for
 $m_h=120\gev$. Moreover, by combinig the event rates in $Z+h$ and weak boson
 fusion, they noted, 
 the Higgs boson mass could be extracted from the production cross section. Another possibility
to discover a Higgs that decays into neutralino pairs is to rely on the remaining
visible channels. But even them are indirectly affected by the decay $\hnn$.

\EPSFIGURE[t]{mub,scale=0.4}{BR($h\to \gamma \gamma$) as a function of the
  neutralino mass for $\tan\beta=3$ and different values of $\mu$.\label{fig:mub}}
By increasing the total decay width of the Higgs, the process $\hnn$
suppresses the branching ratio into \emph{all} other channels, including the
visible ones. If neutralinos are lighter than $m_h/2$, the Higgs branching
ratios into Standard Model particles will be reduced  by the factor $1-$\brhnn
with respect to the conventional models --where neutralinos are heavier than
$m_h/2$. This suppression factor is universal, it equally affects all other
decay modes. 

Figure \ref{fig:mub} shows, as an example, the BR($h\to \gamma\gamma$) as a function of the
neutralino mass for $\tan\beta=3$ and different values of $\mu$. As expected,
for $m_\chi>m_h/2$ -when the decay $\hnn$ is forbidden- BR($h\to
\gamma\gamma$) is simply a constant. For $m_\chi<m_h/2$, on the other
hand, the branching ratio is smaller than this constant value, with a  suppression
factor depending on $m_\chi$ and $\mu$\footnote{Since \brhnn depends also on
  $\tan\beta$, the suppression will depend on $\tan\beta$ too.}. From the
figure we see that the suppression factor could be more than a factor of
four. If observed, such suppression would provide compelling evidence for these models.

\section{Conclusions}
We have considered light neutralinos in the MSSM and
showed that they are a relevant decay channel of the Higgs boson. The region
in the supersymmetric  parameter space where the Higgs  decay  into
neutralinos is
important was clearly identified: low $\tan\beta$, not so large $\mu$, and
$M_1<50\gev$. After studying \brhnn as a function of these three parameters, we found that it can be
as large as $80\%$. That is, the decay $\hnn$ might be the dominant decay mode
of the Higgs boson. We also pointed out  that, as a result of this additional channel, visible decay modes,
such as $h\to \gamma\gamma$,
are suppressed by up to a factor four.

\end{document}